# Task Scheduling in Cloud Computing Using Hybrid Meta-heuristic: A Review

Sandeep Kumar Patel[1]                                                                 Avtar Singh[2]

[1]*Department of Computer Science and Engineering, Dr B R Ambedkar National Institute of Technology Jalandhar (PB),144011, India*
sandeep.pmg61@gmail.com
[2]*Department of Computer Science and Engineering, Dr B R Ambedkar National Institute of Technology Jalandhar (PB), 144011, India*
avtars@nitj.ac.in

**Abstract.** In recent years with the advent of high bandwidth internet access availability, the cloud computing applications have boomed. With more and more applications being run over the cloud and an increase in the overall user base of the different cloud platforms, the need for highly efficient job scheduling techniques has also increased. The task of a conventional job scheduling algorithm is to determine a sequence of execution for the jobs, which uses the least resources like time, processing, memory, etc. Generally, the user requires more services and very high efficiency. An efficient scheduling technique helps in proper utilization of the resources. In this research realm, the hybrid meta-heuristic algorithms have proven to be very effective in optimizing the task scheduling by providing better cost efficiency than when singly employed. This study presents a systematic and extensive analysis of task scheduling techniques in cloud computing using the various hybrid variants of meta-heuristic methods, like Genetic Algorithm, Tabu Search, Harmony Search, Artificial Bee Colony, Particle Swarm Optimization, etc. In this research review, a separate section discusses the use of various performance evaluation metrics throughout the literature.

## 1      Introduction

Cloud computing is an on-demand availability of shared resources i.e. storage, computation power, network, software, and other services to fulfill client requests in small time and cost over the internet. The advantages include resource-transparency, reliability, affordability, flexibility, location –independence and a high availability of services [1]. To achieve these functionalities, a proper task scheduling is required so that it can provide a good performance in a swift manner. Moreover, cloud computing aims to satisfy the customer requirements in view of the Service Level Agreement(SLA) and the Quality of Service(QoS) [2]. There exist basically three service models viz.; 1. Platform as a service(PaaS), 2. Infrastructure as a service(IaaS) and 3. Software as a service(SaaS) which can be deployed on various deployment models like Private Clouds, Public Clouds and Hybrid Clouds. [3].                             Virtualization allows sharing a single instance of a resource among multiple people e.g. server, network, desktop, operating system. It is used to display the hallucination, rather than actual, of many isolated virtual machines. Each VM runs many guest operating systems to ensure the heterogeneity of application. In this scenario, Hypervisor plays a major role as it assists the interaction between guest OS and physical hardware [4].

A key concept in cloud computing is the Resource Management which is implemented in two stages. The first stage- Resource Provisioning, provides means for the selection, deployment and management of software from task submission to task execution as requested by an application. The second stage- Task Scheduling, is the process of mapping of various incoming tasks to existing resources to achieve an optimal execution time and an efficient resource utilization [5]. The total completion cost of any task is the summation of the communication cost and execution cost of that task. The Data transfer cost may also be considered for large data transfers. To minimize this cost, resources are equally distributed among the tasks.

In this research area, numerous studies have been done over the years with meta-heuristic techniques being the most prevalent in the literature. Poonam et. al. [6] present a summarized study of various meta-heuristic optimization techniques employed in the cloud computing environment for task scheduling. Our study is the only one that focuses on hybrid techniques. The primary objective of the study is to conduct a proper systematic comparative analysis of various hybrid distinctions based on the metrics like makespan, cost, throughput and energy consumption. Aiming to infer intrinsic behavioral properties to these algorithms and assist in the appropriate and efficient hybridization. To build a roadmap for future studies is the ultimate outcome of this research.

The organization for the rest of the paper is as follows: section 2 presents a brief description of the task scheduling in the cloud environment. In section 3, various optimization techniques are discussed. In section 4, a literature review of hybrid meta-heuristic techniques for scheduling is presented. Section 5 and 6 gives a tabular summary of the related works and comparison of performance metrics respectively and the conclusion and the future work are presented in section 7.

## 2      Task Scheduling In Cloud

The task scheduling in cloud environment is an NP-complete problem so it is hard to find an optimal solution in polynomial time. The scheduling in cloud improves the resource utilization and reduces the overall completion time. There does not exist a standard task scheduling technique that could be extended to a large-scale environment. The main job of task scheduler is to distribute customer requests to all the present resources to execute them. Task scheduling becomes very important from the user's point of view as they have to pay based on usage of resources based upon time. There are different effective resource scheduling criteria which reduces execution cost, time, energy and increases CPU utilization and productivity. A broad classification can be done into the following categories:  static, dynamic, preemptive, non-preemptive, centralized and decentralized scheduling [7]. The major performance metrics used in the literature are as follows [8][9]:

- The *Makespan* is the maximum finishing amongst all the received tasks.

$$\text{Makespan} = \text{Max}\{ FT_i \mid \forall_i \} \qquad (1)$$

- The *Throughput* is the number of tasks completed with respect to deadline of each job.

$$Throughput = \sum_{i \in I} X_i \qquad (2)$$

- The *Response Time* is the time at which task arrives in the system to the time task is scheduled first time for execution.

$$Response\ Time = T_{first\ execution} - T_{arrival} \qquad (3)$$

- The *Transmission Time* is time required to transfer a task from queue to a specific VM.
- The *waiting time* is defined as the time consumed in the waiting queue before the start of execution of particular task.
- The *Total Cost* depends on transfer of file and processing time.

$$Total\ Cost = P_i \times P_c + \left\{ \sum_{f \in FIN_i} Size(f) + \sum_{f \in FOUT_i} Size(f) \right\} \times PTPB \qquad (4)$$

Where $P_C$ processing cost, f is file, PRTB processing time per bytes.

## 3      Optimization Techniques

The performance of a system is directly influenced by the efficiency of task execution schedule. To achieve this, a number of optimization algorithms for allocating and scheduling the resources proficiently in the cloud have been proposed over the years. A comparative study of different meta-heuristic techniques is presented here that perform efficient task scheduling is given below:

### 3.1    Genetic Algorithm (GA)

It was introduced by Holland in 1975. It is inspired from the biological idea of creating new generation population. Like in the Darwin's theory of Natural Selection, the term "Survival of the fittest" is employed as the strategy method for task scheduling as the tasks are assigned to resources according to the value of fitness function. The basic terminologies of the GA are defined below [10-11]:

- *Initial Population:* This is defined as the set of all solutions(individuals) that are used by GA to find out the optimal solution.
- *Fitness Function:* The fitness value specifies the productivity of a solution (individual). It is the measure of the fitness of the existing individual (solution) in the population.
- *Selection:* The selection technique is used to choose a solution for the improvement to generate the next generation population. This operation drives the GA based on fitness. The various selection techniques like: roulette wheel, tournament selection, and rank based selection.
- *Crossover:* This is done by selecting two parent solutions and then generating a new solution tree by intermixing the parts of those parents.
- *Mutation:* It is an operator that produces genetic diversity in the population. It alters one or more parts of the solution from its initial state. This can introduce entirely new gene (solution constituents) in the population.

### 3.2 Harmony Search Algorithm (HS)

HS is a meta-heuristic search algorithm inspired from the process of musicians searching for a perfect harmony [12]. The main principles in the HS are described as follows:
- *Initialization:* Initialization of harmony search parameters like Harmony Memory Size (HMS), Pitch Adjusting Rate (PAR) and Harmony Memory Considering Rate (HMCR).
- *Initialize the Harmony Memory (HM):* In this step harmony memory HM (a 2-D matrix containing a set of possible solutions) is randomly initialized.
- *Improvise a new harmony:* The generation of a new harmony is known as improvisation. To create a new solution(harmony) following three rules used:
    - Exactly the same from memory.
    - Similar to known one after pitch adjustment.
    - Totally compose new one.

$$HMold = HMnew + Bandwidth \ X \ RandGen \qquad (5)$$

where, $RandGen$ (-1,1)

- *Randomization*: It increases the diversity of solution.
- *Harmony memory updation:* For each harmony the value of objective function is computed. Then, the new harmony vector is compared with the previous one. If the new harmony vector is better than the worst harmony in the HM, it takes place of the worst.

### 3.3 Tabu Search (TS)

TS is a meta heuristic optimization search algorithm which uses a memory like HS. Tabu search was proposed by Glover [13]. It starts with single random solution and is updated by one of the neighboring solutions. This process continues until the most optimal solution is found. The main principles of TS are as follows:
- *Initial Solution:* The initial solution is found by a greedy heuristic method.
- *Initialize Tabu List:* Tabu search generates a neighborhood solution from the current solution and accepts a solution as the best solution if it is not improving the previous solution. This method can form a cycle by regenerating a previous solution again. Hence to avoid this cycle, TS discards the previous visited solution using memory called Tabu list.
- *Fitness function:* The fitness function chooses the best solution. When best solution is found, it is kept in memory otherwise it is removed from memory or tabu list.
- *Updation:* Update Tabu list S=S', where S is the previous solution and S' is the newly created better solution.

### 3.4 Particle Swarm Optimization (PSO):

PSO has recently become an important heuristic approach and has been applied to various computationally hard and complex problems, such as task scheduling problem, extraction of knowledge in data mining, electrical power systems, etc. It draws inspiration from social behavior of organisms like a bird flock or fish schooling. The main principles of the PSO are defined below [14]:

- *Initial population:* This is defined as the set of all solutions that are randomly generated in search of an optimal solution. The solutions in the population are termed as the Particles.
- *Fitness Function:* The fitness value is responsible for the productivity of any particle. It measures the fitness of an existing individual (solution) in the population.
- *Selection:* In each iteration there are two parameters that are responsible for determining the next position of each particle: the personal best (p-best), which the individual particle has during its exploration; and S the global best (g-best), which is the best position that a particle has among all the particles.
- *Updation:* After calculating these two best values, the updation of velocity and position of a particle is done using the following equation:

$$Vnew = Vold + c1 \; X \; rand() X (pbest - present) + c2 \; X \; rand() X (gbest - present) \quad (6)$$

$$Present \; new = present \; old + Vnew \quad (7)$$

where V is the particle velocity, c1, c2 are the learning factors and rand() is a random number between (0, 1).

### 3.5 Cuckoo Optimization Algorithm (COA):

It is inspired from obligate brood parasitism of cuckoo which lays eggs in the host nest having Lévy flight behavior of the birds [15]. The main terminologies of this algorithm are [12]:

- *Initialization:* It is an initial population of solutions, $S_x$, which is created randomly, where x = 1,2,…n.
- *New Cuckoo Generation*: In this step, new cuckoos(solutions) are generated using levy flights.
- *Fitness Evaluation*: Once a solution is generated, its fitness is calculated and the best one is selected.
- *Updation*: The new solution is created using the equation:

$$Inew = Iold + \alpha \; Levy(\lambda) \quad (8)$$

- *Selection/Rejection:* The solution which has worst fitness value is thrown out of solution space.

### 3.6 Artificial Bee Colony (ABC):

ABC algorithm is a swarm based meta-heuristic optimization technique, inspired from foraging conduct of honey bee colonies. ABC algorithm classifies the bees into three types: employed bees, scout bees and onlooker bees. The employed bees search for the food around the food source in their memory and this info of food sources is passed on to the onlooker bees. The on looker do the selection procedure from the food sources found by the employed bees. The probability of selection of a food source by the onlooker bees is determined by its quality. The scout bees induct the diversity by abandoning their food sources and getting along in search of new ones. The total number of employed bees or the onlooker bees is the total number of solutions in the swarm [16]. The main phases of ABC Algorithm are:

- *Initialization Phase:* It is a randomly initialized initial population of $S_N$ solutions (sources of food), where N represents the swarm size.
- *Employed Bee Phase*: It determines the neighborhood food source, denoted by Vm. The Fitness of each food source is also calculated in this phase.

$$Vi,j = Xi,j + \emptyset(Xi,j - Xk,j) \tag{9}$$

- *Onlooker Bee Phase:* The quality of a food source is estimated by its profitability and the effectiveness of all food sources.
- *Scout Bee Phase*: The new solutions are randomly searched by the scout bees.
- *Fitness value:* The fitness function is used to choose the best solution.

### 3.7 Ant Colony Optimization (ACO):

ACO approach was introduced by Dorigo in 1992. It is a meta-heuristic method inspired from food searching method of ants. The ants share the food source information through pheromone path. An ant solves a problem by using a construction graph where edges are the possible partial solution that the ant can take according to a probabilistic state transition rule. After selection of either a partial or a complete solution, the pheromone updating begins to start. This rule gives a mechanism for speeding up convergence and also prevents premature solution stagnation [17-18].

### 3.8 Simulated Annealing (SA):

Simulated Annealing is an iterative meta-heuristic random search optimization technique for solving nonlinear optimization problem. The name and motivation originate from annealing in metallurgy, a process of heating and controlled cooling of a material to increase the size of its crystal and diminish their defects. It was proposed as the metropolis algorithm and after that many variations were introduced later on. Simulated annealing is widely being used in task scheduling in cloud environment, machine -scheduling and vehicle routing etc. [19].

### 3.9 Bacteria Foraging Optimization Algorithm (BFO):

BFO was proposed by Kevin Passino (2002) and includes three basic mechanisms: chemotaxis, reproduction, and elimination-dispersal. Chemotaxis helps the movement of E-coli cell by swaying and plummeting with help of flagella. Reproduction: Only half of population survives, and that bacterium degenerates into two identical ones, which are then positioned at the same location leaving the total bacteria population unaffected. Elimination and Dispersal: The chemotaxis is considered for local search and it increases rate of the convergence. Since bacteria can get stuck in local minima hence, the diversity of BFO is changed to disregard the chances of getting stuck in the local minima. The event of dispersion occurs after a particular number of reproduction processes. So, some bacteria are taken with probability *P*, to be killed and shifted to a different location within the environment.[20]

### 3.10 Gravitational Search Algorithm (GSA):

GSA is an optimization technique method based on "Gravitational Law" [21]. This algorithm is basically population based multi-dimensional optimization algorithm where agents are called as objects and their performance can be calculated by their masses. The masses are the way of communication as the agents move towards heavier masses by gravitational force. The heavy masses correspond to good solutions and move slowly than lighter ones. Each agent (mass) has four characteristics: position, Active Gravitational Mass (AGM), inertial mass(IM), and Passive Gravitational Mass(PGM). The solution of the problem can be obtained by position, and its inertial masses and gravitational can be calculated using a fitness function.
- *Initialization:* All the agents are initialized with a different mass.
- *Fitness Function:* Here, It is used to calculate the masses

$$Mi = \frac{fitt(t) - worstt(t)}{bestt(t) - worstt(t)} \tag{10}$$

where $fitt(t)$, $bestt(t)$, $worstt(t)$ denote fitness of Xi, $\min\{fitt(t)\}$, $\max\{fitt(t)\}$

- *Evaluate Force:* At time 't', the force acting on mass 'i' from mass 'j' defined as:

$$F(t) = G(t) \frac{Mpx(t) \times May(t)}{Rxy(t) + e} (Rxy(t)) \quad (11)$$

where M$ax$ is the AGM, Mpx is the PGM, G(t) is the universal gravitational constant at time t, e is a constant, and Rxy(t) is the Euclidian distance between agents i and j:

- *update position and velocity:*

$$vi(t+1) = randi \times vi(t+1) + ai(t) \quad (12)$$

$$X(t+1) = X(t) + V(t+1) \quad (13)$$

### 3.11 Lion Optimization Algorithm (LOA):

It is a meta-heuristic algorithm inspired from the lion. The lion has two types of social organization: resident and nomad. Residents lives in groups, called pride that includes one or more than one adult males, around five females and their cubs. The nomads move about sporadically either in pairs or single. A lion can switch lifestyle means nomads may become residents and vice –versa [22].

- *Initialization:* A random solution(Lions) set is initialized.
- *Fitness Value:* It improves the solution.
- *Calculate nomads and prides:* A part of population is randomly chosen as prides and rest of population is selected as nomads.
- *Hunting:* The food requirement of the pride is collected by a group of females. And hence they apply certain strategies to trap and capture the prey. Each female changes its location based on the relative location of the other members of the pride.
- *Roaming:* Every male lion roams around in the territory of that pride. A resident male's best solution is updated if it visits a place which is better than its personal best.
- *Mating:* It is an essential process that ensures the survival of the lion and assists in the information exchange with other members. A female and a randomly selected male from the same pride produce an offspring.
- *Defense:* A mature male lion becomes aggressive and contests other males in its pride. Once the lion is beaten, it abandons its pride and becomes a nomad.

## Hybrid Meta heuristic approaches:

Every meta-heuristic algorithm comes with its share of pros and cons. Clubbing together a selected set of them to harness the advantages of each one can improve the efficiency. Several such hybrid approaches have been proposed in the literature, which have been discussed below:

### 3.11 The Harmony Tabu Search (THTS):

In this proposed method, TS and HS is combined to improve the result. TS is applied in first step followed by the HS. At the beginning of the algorithm, TS is initialized with a tabu list that contain all the candidate solutions and generates initial solutions which are compared with the best candidate solution in the tabu list. Its better quality guarantees its inclusion into the tabu list. After this, HS is applied with initialization of Harmony memory (HM) with the tabu list. A new solution is obtained from HM by improvising each components of solution with harmony memory considering rate (HMCR) parameter and mutation of the solution by pitch adjusting rate (PAR) [23].

**3.12 Cuckoo Harmony Search Algorithm (CHSA):**

The CS is very efficient for local search with a single parameter. But it has a limitation that it takes huge amount of time to obtain an optimal solution. Similarly, HS has a limitation too, its search execution completely depends upon the parameter setting. When hybridization is applied, it is seen that it removes those limitation which affect the performance of CS and HS individually [12].

**3.13 Harmony-Inspired Genetic Algorithm (HIGA):**

This Hybrid algorithm is composed of the HS and Genetic algorithm to detect both local optima as well as global optimal when task scheduling is being done. The HIGA provides better results when a scenario arises where the best individual remains in the same state either in local optimal state or global optimal state after many generations with the help of HS and updates the current population in the GA. If HS failed to find it in much iterations, it simply means the best solution might be in global optimal state. As a result, process can halt. So, in spite of halting process, The HIGA algorithm reduces the number of iterations and senses local or global optimal state every time. In this, GA is considered as primary optimization algorithm and when local optimal solution is found by any individual then HS is used to find global optimal solution [24].

**3.14 Genetic Algorithm Particle Swarm Optimization (GA-PSO):**

Here the GA is applied first and a random population is generated. Then fitness function is applied to obtain elites which are divided into two halves. First half part is enhanced by GA and another half by PSO. In GA, the best elites are given to crossover operator and mutation operator, while in PSO pbest and gbest is calculated for each elite. The position and velocity of elites is calculated and updated in each iteration [25].

**3.15 Multi-Objective Hybrid Bacteria Foraging Algorithm (MHBFA):**
This algorithm produces a solution with a better local and global search capability and a greater convergence time. Since, Bacteria Foraging (BF) has a great local search capability and unluckily has a poor global search. GA overcomes this limitation hence, the MHBFA inherits swarming, elimination and dispersal from BF and these are measures which are critical in global search procedure [26].

**3.16. Simulated Annealing based Symbiotic Organisms Search (SASOS):**

This Hybrid algorithm is comprised of the Simulated Annealing (SA) and Symbiotic Organism Search (SOS) for achieving the improved convergence rate and improved quality of the solution. The SOS algorithm includes phases like mutualism, commensalism, and parasitism. The SA has a systematic ability to get better local search solutions using the procedure of commensalism and mutualism phases of the SOS. The parasitism phase remains unaffected because it deletes the passive solutions and injects the active ones in the solution space which could help the search process out of the local optimal region [27].

**3.17. The Technique for Order of Preference by Similarity to Ideal Solution-Particle Swarm Optimization (TOPSIS-PSO):**
In this hybridization technique, PSO is combined with TOP-SIS algorithm to find an optimal solution by taking into account criteria like, execution time, transmission time and cost, which is carried out in two phases. In the first phase, TOPSIS is applied in order to get the relative closeness of the jobs. In the second phase, PSO is applied for all tasks to compute closeness in these three criteria in all virtual machines (VM). The fitness function of PSO is formulated using TOPSIS which gives an optimal solution in minimum time [28].

### 3.18. Artificial Bee Colony Simulated Annealing (ABC-SA):

This Hybrid algorithm is comprised of ABC and Simulated SA for the efficient task scheduling depending upon their sizes, priority of request came etc. [29].
.

### 3.19. Genetic Algorithm Artificial Bee Colony (GA-ABC):
This Hybrid algorithm combines the features of GA and ABC with the facility of Dynamic voltage and frequency scaling (DVFS) to achieve efficient task scheduling. In this algorithm, GA is used as first step for starting allocation process of tasks to VM and obtained the new individuals until termination condition of GA occurs. The output of GA is fed as the input to the ABC. Then, ABC provides the optimal distance between task and VMs [30].

### 3.20. Cuckoo Gravitational Search Algorithm (CGSA):
This hybrid CGSA composed of CS and GSA. The major demerit of CS algorithm is that it takes maximum time in order to find the optimal solution and the disadvantage of GSA is that it does not converge well for local optimal solution. The CGSA uses the advantages of CS and GSA It conquers the weaknesses and provides the efficient solution in a shorter computational time [31].

### 3.21. Oppositional Lion optimization algorithm (OLOA):
This hybrid OLOA uses the benefits of Lion optimization algorithm (LOA) and oppositional based learning (OBL).In this hybrid approach, OBL is nested within the LOA [32].

### 3.22. Fuzzy system and Modified Particle Swarm Optimization (FMPSO):
PSO uses the Shortest Job to Fastest Processor (SJFP) technique to initiate the initial population, position matrix of particle an velocity matrix. The roulette wheel selection, crossover operator and mutation operators are considered to overcome the drawbacks of PSO like the local optima. The Hierarchical fuzzy system is used for the evaluation purpose of fitness value of each particle [33].

## 4   Literature Review

The related studies on this research area have been discussed in Table 1. Hadeel Alazzam et al.[23] proposed a hybrid task scheduling algorithm which includes Tabu- Harmony search algorithm(THTS). The algorithm performs better in respect of makespan and cost compared to TS, HS, round-robin individually. K.Pradeep et al.[12]presented hybrid Cuckoo Harmony Search Algorithm (CHSA)for task scheduling to improve the energy consumption, memory usage, credit, cost, fitness function and penalty and it was observed that the performance of this proposed algorithm is comparatively better than individual CS and HS algorithm, and hybrid CSGA. Mohan Sharma et al.[24] focused on a Harmony Inspired Genetic Algorithm(HIGA)for energy efficient task scheduling to improve energy efficiency and performance. The results describe that the presented algorithm improved efficiency and performance. A. M. Senthil Kumar et al.[25] discussed a hybrid Genetic Algorithm Particle Swarm Optimization (GA-PSO) to minimize the total execution cost. GA-PSO helped to obtain the result better than various existing algorithms like GA, Max-Min, Min-Min.

Sobhanayak Srichandan et al.[26] discussed a Hybrid Bacteria Foraging Algorithm (HBFA) for task scheduling which inherits the desirable characteristics of GA and Bacteria foraging (BF) in cloud to minimize the makespan and reduce energy consumption economically as well as ecologically. The results show that HBFA outperforms than GA, PSO, BF when applied alone. Mohammed Abdullahi et al.[27] put forth a hybrid algorithm to optimize the task scheduling based on SA and SOS for improving convergence speed, response time, degree of imbalance and makespan. The results show that SASOS performs better than SOS. Neelam Panwar et al.[28] proposed a new hybrid algorithm based on TOPSIS and PSO to solve multiple objective such as transmission time, resource utilization, execution time, and cost. The achievement of TOPSIS PSO has been compared with ABC, PSO, dynamic PSO (DPSO), FUGE and IABC algorithm in terms of transmission time, makespan, resource utilization and total cost.

Muthulakshmi et al.[29] proposed a hybrid algorithm which combines the advantages of ABC and SA to improve the makespan. The result obtained by using this algorithm outperforms than MFCFS, Shortest Job First(SJF), LJF, hybrid ABC-LJF and hybrid ABC-SJF. Sunil Kumar et al.[30] has presented a hybrid algorithm GA-ABC to make improvement in makespan and energy consumption using DVFS. DVFS model is used for the calculation of power consumption. The results show better results than Modified Genetic Algorithm (MGA).K. Pradeep et al.[31] discussed a hybrid algorithm which inherits the benefits of both Cuckoo Search (CS) and Gravitational Search (GS) to execute the tasks with low cost, less usage of resources, and minimum energy- consumption. The results show that CGSA perform better than CS,GSA, GA, PSO.Pradeep Krishnadoss et al.[32] presented a hybrid algorithm that uses LOA and Oppositional Based Learning(OBL) to improve makespan and cost. The OLOA performs better than PSO and GA.Ben Alla et al. [33] proposed two hybrid algorithms using Fuzzy Logic with PSO and SA with PSO for optimization of makespan, waiting time, cost, resource utilization, degree of imbalance and queue length of the tasks in cloud environment. The hybrid algorithm outstrips the individual SA and PSO in their performance.

Rasha A. Al-Arasi et al.[34] presented hybrid algorithm that inherits the advantages of GA with Tournament selection and PSO. The GA-PSO provides better results by reducing makespan and increasing the resource utilization. A. Kousalya et al.[35] implemented a hybrid algorithm that uses improved GA including divisible task scheduling into the foreground and background process and PSO. The GA PSO performs better in terms of execution time and resource utilization. Bappaditya Jana et al.[36] presented a hybrid GAPSO algorithm to provide comparatively better response time and minimize the waiting time. The results show that this cost-effective GA PSO achieves better response time, and minimizes the waiting time.GAN Guo-ning et al.[37] discussed about hybrid algorithm using GA and SA which considers the Quality of Service (QOS) requirements for many types of tasks, that correspond to the user's tasks-characteristics in cloud –computing environment. Hua Jiang et al.[38] focused on hybridization using merits of HS and SA which provides global search and faster convergence speed and local minima escaping to get the better solutions. Medhat A. Tawfeek et al.[39] proposed a hybrid swarm intelligence technique which involves ABC, PSO, ACO. The algorithm performs better than existing algorithms.

Najme Mansouri et al.[40] presented a hybrid algorithm FMPSO to determine the execution time, makespan, imbalance degree, improvement ratio and efficiency,. The results show that it does better than other strategies like FUGE,SGA, MGA etc. Poopak Azad et al. [41] discussed a hybrid algorithm based on Cultural Algorithm which considers acceptance and influence as major operators and the Ant Colony Optimization Algorithm minimizes the makespan and energy consumption. The results show that it performs better than HEFT and ACO. Jun-qing Li et al. [42] focused on a hybrid task scheduling technique with ABC algorithm with flow shop scheduling for improvement of convergence rate.

Hicham N. Manikandan et al.[43] proposed a hybrid algorithm uses the benefits of LOA and GSA for the multi-objective task scheduling and uses profit, cost, and energy as the performance metrics. The LGSA perform better than the others. Danlami Gabi et al.[44] presented a hybrid multi-objective algorithm comprised of Cat Swarm optimization (CSO) and SA for task scheduling. The algorithm outperformed it constituents by resulting in minimum execution time, cost and a greater scalability which provides global search and faster convergence speed and local minima escaping to get the better solutions.

**Table 1.** Literature Survey Summary.

| Authors | Strategy | Description | Performance Metrics Used | Achievements | Environment |
|---|---|---|---|---|---|
| Hadeel Alazzam et al.(2019) | The Harmony Tabu Search (THTS) | Tabu search is applied as first step after that Harmony search is applied until optimal solution obtained. | • Makespan<br>• Throughput,<br>• Total cost | • Least makespan<br>• Least cost<br>• Nearly the same throughput. | CloudSim |
| K. Pradeep et al. (2018) | Cuckoo Harmony Search Algorithm (CHSA) | Local optimal solution with cuckoo search and then it is given to Harmony Search | • Memory usage<br>• Cost<br>• Energy consumption<br>• Fitness<br>• Penalty<br>• Credit | • Memory usage of 0.156<br>• Cost of 0.0098$.<br>• Energy consumption is 0.23<br>• Minimum Fitness function gives the high profit.<br>• Penalty value is 0.276<br>• Credit is 0.724 | CloudSim |
| Mohan Sharma et al.(2019) | HIGA | Current generation is evolved by GA results in local optima and it is given to HS through which global optima is achieved. | • Makespan,<br>• Energy consumption,<br>• Execution time | • Makespan improved by 47%<br>• Energy saving 33%,<br>• Less execution time by 39% | MATLAB 2013b |
| A. M. Senthil Kumar et al.(2019) | GA-PSO | The population is randomly generated. The half population is evaluated GA and rest half is by PSO and combine both result to get optimal solution. | • Response time | • Response time is lowered by 1678, 1393, and 1000 ms comparison done with Min-Min, Max-Min and GA. | CloudSim |
| Rasha A. Al-Arasi et al. (2018) | HTSCC | GA is employed to randomly generated solution with tournament selection operator, If optimal solution is find then it is given to PSO to find best solution | • Makespan,<br>• Resource utilization | • Makespan 31.32% and 22.36%,<br>• Resource utilization 23.17% and 19.6% better than GA and PSO | CloudSim |
| A. Kousalya et al. (2017) | Improved GA-PSO | Group of PSO is ordered and coefficient of every constraint of optimal solution of GA is evaluated. Best solution is obtained if anyone meets termination criteria. | • Cost<br>• Execution time | • Least cost<br>• Least execution time | CloudSim |
| Bappaditya Jana et al. (2018) | Enhanced GA - PSO | Implemented with GA and obtained result is given to PSO to achieve best result | • Response time<br>• Waiting time | • Minimum waiting time<br>• Minimum response time | CloudSim |

| Author | Algorithm | Description | Parameters | Results | Tool |
|---|---|---|---|---|---|
| SobhanayakSrichandan et al. (2018) | MHBFA | Initialized with BF and Hybrid chemotaxis and Hybrid reproduction is employed by PSO to obtain optimal solution. | • Makespan, • Energy consumption • Convergence •Stability, \ • Solution diversity. | • Makespan decreases, • Minimum energy consumption thanGA, PSO BFA • Scalable • High mean coverage ratio | MATLAB 2013b |
| Mohammed Abdullahi et al. (2016) | SASOS | SOS is initialized first and SA techniques is applied during mutualism and commensalism phase of SOS. | • Convergence speed, • Response time, • Makespan. | • Faster convergence for 500 tasks •.Improved response time • Least makespan | CloudSim |
| Neelam Panwar et al. (2019) | TOPSIS –PSO | Initialized with PSO in which fitness is calculated using TOPSIS. | • Makespan, • Transmission time, • Cost • Resource utilization | • Least makespan • Minimum transmission time, • Cost • Improved resource utilization | CloudSim |
| Sunil Kumar et al. (2019) | GA-ABC | Started with GA and when termination occurred then result is given as input ABC followed by initialization of DVFS. | • Makespan • Energy consumption. | • Less makespan by 75.5% for 40 tasks • Less energy consumption by 84.14% for 40 tasks than Modified GA | CloudSim |
| B. Muthulakshmi et al. (2017) | ABC-SA | Initialized with ABC having random selection capability of SA to increase efficiency. | • Task size, • Priority of the request | • Improved task size and priority of request | CloudSim |
| Pradeep Krishnadoss et al. (2018) | OCSA | Initialized with OBL and fittest solution is selected and updating is done by CS. | • Makespan • Cost | • Makespan value is 141.5 for 500 tasks • cost is 110.3 for 500 tasks | CloudSim |
| GAN Guo-Ning et al. | GSAA | Population employed with GA. After mutation phase the result is given to SA. | • Bandwidth • Completion time • Cost • Reliability as QoS parameter | • Converges in 743 generation • function-value is 0.91174 • Optimal Scheduling is achieved. | MapReduce |
| Hua Jiang et al. (2012) | HSSA | Employed SA to population generated using partial HS | • Completion time | • Least Completion time. | VC++6.0 |

| Medhat A. Tawfeek et al. (2016) | Hybrid swarm intelligence techniques | ABC is first initialized and the solution is handled by different suitable module like Bees(), Ants(), particles () etc. | • Makespan | • Least makespan than ABC, PSO, ACO | CloudSim |
|---|---|---|---|---|---|
| Najme Mansouri et al. (2019) | FMPSO | Initialized with PSO in which fitness of solution is calculated using fuzzy interference system. | • Makespan<br>• Improvement ratio<br>• Efficiency<br>• Execution Time (ET). | • Makespan by 13% in comparison with FUGE<br>• Reduce ET by 8% and 16%<br>• Average efficiency of 3.36. | CloudSim |
| Poopak Azad et al. (2017) | HCACO | Initialization is done with ACO and the obtained local result is given to cultural Algorithm | • Makes span,<br>• Energy conservation | • Completion time 106.48<br>• Energy consumption .204 | C# language in cloud azure |
| Jun-qing Li et al. (2019) | ABC-HFS | Employed with ABC having 2 kinds; 1.HFS with identical parallel machines and 2.HFS with unrelated machines. | • Completion time | • Reduced completion time | C++ |
| Pradeep Krishnadoss et al. (2019) | OLOA | LOA is implemented with initialization of population based on OBL | • Makespan<br>• Cost | • Makespan 95.2 sec.<br>• cost is 65.2 sec | CloudSim |
| N. Manikandan et al. (2019) | LGSA | Implemented with LOA and fitness function is evaluated using GSA | • Profit,<br>• Cost,<br>• Energy. | • Max profit 0.8<br>• Min cost 0.011$<br>• Min energy 0.039 | CloudSim |
| Hicham Ben Alla et al. (2018) | Dynamic Queue Meta-heuristic Algorithm | The FL-PSO algorithm and SA-PSO are applied to get the optimal solution | • Waiting time,<br>• Makespan,<br>• Cost,<br>• Resource utilization | FL-PSO gives the improved results of<br>• Wating time, •. Makespan,<br>• Cost,<br>• Resource utilization | CloudSim |
| Danlami Gabi et al. (2018) | CSM-SA | CSM is used as a first step then SA is implemented as second step. | •Time<br>•Cost | Improved<br>•Time<br>•Cost | CloudSim |

## 5  Comparison of Performance Metrics

The selection of appropriate performance evaluation metrics is also important in determining the efficiency of a scheduling algorithm. There have been numerous metrics devised over the years to capture the overall efficiency of the algorithm. Achieving that with a single metric is not possible, making the use of multiple metrics for the evaluation of an algorithm a common trend in the literature. Table.2 presents a summary of the metrics used by each author throughout the literature. Fig.1 is the graphical depiction of Table 1. i.e the number of metrics used by several authors in the literature. The most commonly used metric in the literature is the makespan which can be seen in the Fig.2.

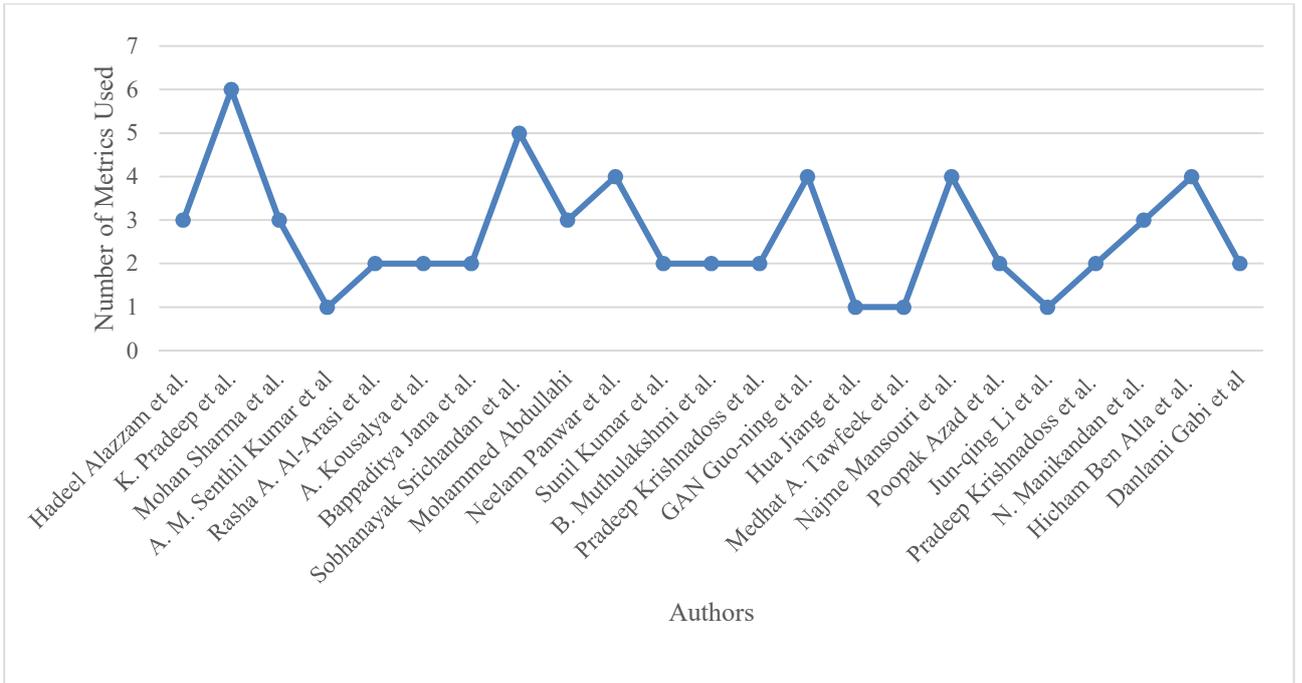

**Fig.1.** Comparison on the basis of metrics used.

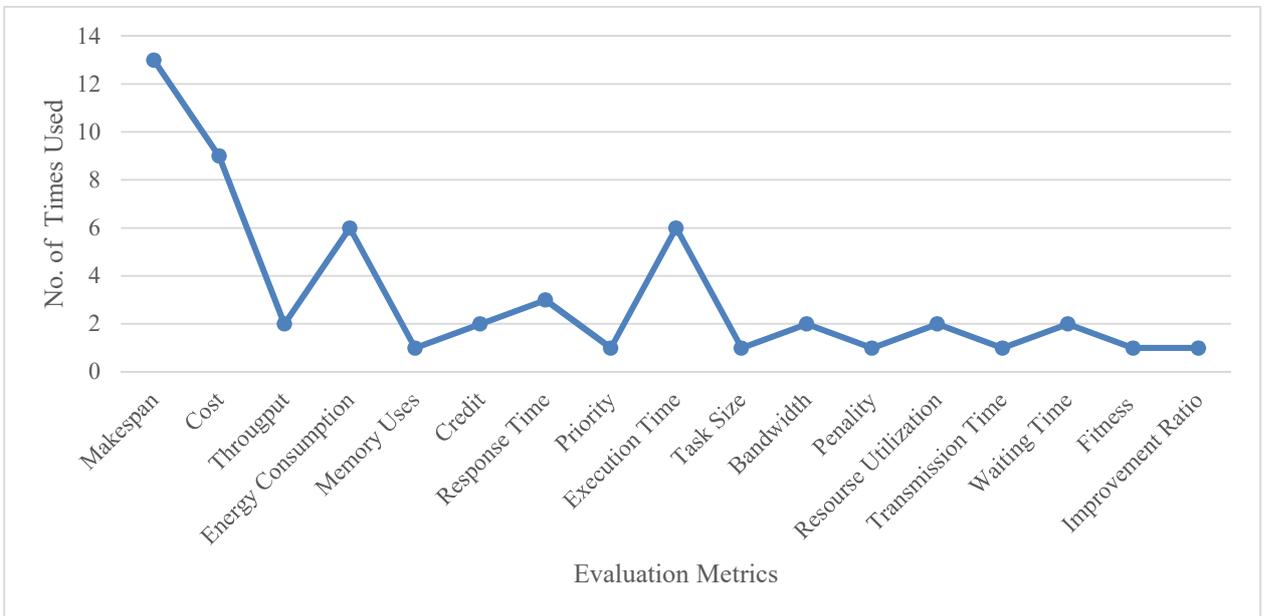

**Fig.2.** Comparison of use of different evaluation metrics.

**Table 2.** Performance Metrics Used.

| Author | Algorithm | Resource Utilization | Makespan | Throughput | Total Cost | Memory Use | Energy Consumption | Credit | Penalty | Fitness | Response Time | Execution Time | Bandwidth | Waiting Time | Transmission Time | Task Size | Priority | Improvement Ratio |
|---|---|---|---|---|---|---|---|---|---|---|---|---|---|---|---|---|---|---|
| Hadeel Alazzam et al. | THTS |  | ✓ | ✓ | ✓ |  |  |  |  |  |  |  |  |  |  |  |  |  |
| K. Pradeep et al. | CHSA |  |  |  | ✓ | ✓ | ✓ | ✓ | ✓ | ✓ |  |  |  |  |  |  |  |  |
| Mohan Sharma et al. | HIGA |  | ✓ |  |  |  | ✓ |  |  |  |  |  |  |  |  |  |  |  |
| A. M. Senthil Kumar et al | GAPSO |  |  |  |  |  |  |  |  |  |  | ✓ |  |  |  |  |  |  |
| Rasha A. Al-Arasi et al. | HTSCC | ✓ | ✓ |  |  |  |  |  |  |  |  |  |  |  |  |  |  |  |
| A. Kousalya et al. | GAPSO |  |  |  |  |  |  |  |  |  |  | ✓ | ✓ |  |  |  |  |  |
| Bappaditya Jana et al. | GA PSO |  |  |  |  |  |  |  |  |  | ✓ |  |  | ✓ |  |  |  |  |
| SobhanayakSrichandan et al. | MHBFA |  | ✓ |  |  |  | ✓ |  |  |  |  |  |  |  |  |  |  |  |
| Mohammed Abdullahi | SASOS |  | ✓ |  |  |  |  |  |  |  | ✓ |  |  |  |  |  |  |  |
| Neelam Panwar et al. | TOPSIS SA | ✓ | ✓ |  | ✓ |  |  |  |  |  |  |  |  |  | ✓ |  |  |  |
| Sunil Kumar et al. | GA ABC |  | ✓ |  |  |  | ✓ |  |  |  |  |  |  |  |  |  |  |  |
| B. Muthulakshmi et al. | ABS-SA |  |  |  |  |  |  |  |  |  |  |  |  |  |  | ✓ | ✓ |  |
| Pradeep Krishnadoss et al. | OCSA |  | ✓ |  | ✓ |  |  |  |  |  |  |  |  |  |  |  |  |  |
| GAN Guo-ning et al. | GSAA |  |  |  | ✓ |  |  |  |  |  |  | ✓ | ✓ |  |  |  |  |  |
| Hua Jiang et al. | HSSA |  |  |  |  |  |  |  |  |  |  | ✓ |  |  |  |  |  |  |
| Medhat A. Tawfeek et al. | Hybrid Swarm |  | ✓ |  |  |  |  |  |  |  |  |  |  |  |  |  |  |  |
| Najme Mansouri et al. | FMPSO |  | ✓ | ✓ |  |  |  |  |  |  |  | ✓ |  |  |  |  |  | ✓ |
| Poopak Azad et al. | HCACO |  | ✓ |  |  |  | ✓ |  |  |  |  |  |  |  |  |  |  |  |
| Jun-qing Li et al. | ABC-HFS |  |  |  |  |  |  |  |  |  |  | ✓ |  |  |  |  |  |  |
| Pradeep Krishnadoss et al. | OLOA |  | ✓ |  | ✓ |  |  |  |  |  |  |  |  |  |  |  |  |  |
| N.Manikandan et al. | LGSA |  |  |  | ✓ |  | ✓ | ✓ |  |  |  |  |  |  |  |  |  |  |
| Hicham Ben Alla et al. | Dynamic Q MHA |  | ✓ |  | ✓ |  |  |  |  |  |  |  |  | ✓ |  |  |  |  |
| Danlami Gabi et al | CSM-SA |  |  |  | ✓ |  |  |  |  |  |  | ✓ |  |  |  |  |  |  |

# 6   Conclusion

The applications of the cloud computing environment have been spiking up since the past couple of decades. With more and more services and applications being shifting to the cloud, the requirement of developing more efficient and faster-driving algorithms viz. task scheduling, resource scheduling algorithms is also growing. Finding an appropriate cost-effective, efficient and competent scheduling algorithm is a tedious task. The scheduling algorithms used in conventional computing systems fail to perform well in a more constrained cloud environment. Relatively new techniques like LOA and ACO in hybrid form have shown promising results by outperforming the others. The performance evaluation metrics do not capture the comprehensive efficiency of the scheduling algorithm. The most widely used metric is the makespan but lately, there has been a shift towards energy-efficient algorithms increasing the use of energy efficiency metric for performance evaluation. All the studies in the literature have used the basic versions of the individual algorithms in the process of hybridization. In the future, the hybridization can be done with the improved variants of these algorithms like improved harmony search, modified PSO, etc to eliminate the implicit limitations of the basic variants.

Though there are numerous standard data sets available that replicate the active cloud scenario but the research needs to be extended to the dynamic scheduling techniques, making it an open research field for the researchers in the future. So far, Meta-heuristics have been performing altogether quite efficiently but as they draw inspiration from many natural or man-made phenomenon making it susceptible to diverging away from the scientific consistency.